\documentclass[pra, twocolumn, superscriptaddress, longbibliography]{revtex4-2}

\usepackage{amsmath, amssymb, amsfonts, amsthm}
\usepackage{graphicx}
\usepackage{bbm} 
\usepackage[colorlinks=true, citecolor=blue, linkcolor=blue, urlcolor=blue]{hyperref}
\usepackage[utf8]{inputenc} 

\newcommand{\id}{\mathbb I}

\newcommand{\Wcov}{\mathcal{W}_{\mathrm{cov}}}

\DeclareMathOperator{\Tr}{Tr}

\newtheorem{theorem}{Theorem}
\newtheorem{lemma}[theorem]{Lemma}
\newtheorem{corollary}[theorem]{Corollary}
\newtheorem{proposition}[theorem]{Proposition}

\newtheorem{definition}{Definition}

\begin{document}

\title{Background-Free Device-Independent Violations of Causal Inequalities}

\author{Issam Ibnouhsein}
\email{issam@computing-matter.org}
\affiliation{Independent Researcher, Aix-en-Provence, France}

\begin{abstract}
The process-matrix framework describes quantum correlations without presupposing a global causal order, yet its standard formulation implicitly relies on background structure through a fixed Choi--Jamio{\l}kowski identification of local input--output spaces. We analyze how such background assumptions can be treated operationally relative to a \emph{fixed} device-independent interface defined by a causal game. We impose \emph{local-frame covariance}, requiring invariance under independent actions of a physical symmetry group $G$ on each laboratory, thereby excluding symmetry-breaking background resources. Covariance induces a representation-theoretic decomposition into symmetry sectors and symmetry-invariant multiplicity subsystems, introducing physical degrees of freedom that lie outside the declared device-independent interface. We then analyze causal-inequality signatures at the level of interface-observable statistics and identify when symmetry-induced, interface-inaccessible degrees of freedom undermine device-independent certification. A certification is called \emph{background-free} if it arises from a locally covariant implementation and does not rely on hidden control mediated by interface-excluded degrees of freedom. We prove that background-free certifications cannot yield device-independent violations of bipartite causal inequalities in the multiplicity-free regime or when all multiplicity subsystems are classical--classical (CC). Such violations necessarily require non-CC multiplicity, with a concrete sufficient route provided by input--output embeddability of an effective process-matrix structure into non-CC blocks. These results delineate which device-independent causal signatures remain certifiable once both symmetry-breaking background structure and interface-level hidden control are excluded.
\end{abstract}

\maketitle

\section{Introduction}

The process-matrix framework~\cite{oreshkov2012} provides an operational formalism for quantum correlations without presupposing a global causal order, opening a broad landscape of quantum-causal possibilities~\cite{chiribella2013,brukner2014}. While this represents a significant step toward a background-independent treatment of causal structure, the standard formulation nevertheless relies on implicit background assumptions that are usually left unexamined in device-independent (DI) analyses of causal inequalities.

This residual background already appears at the kinematical level. The Choi--Jamio{\l}kowski (CJ) isomorphism fixes an identification between linear maps and operators on tensor-product Hilbert spaces. When adopted consistently across laboratories, this choice effectively selects a common representation of local input--output spaces, corresponding in physical implementations to an alignment of local Hilbert-space frames. Quantum reference-frame theory~\cite{bartlett2007,gour2008,marvian2014,giacomini2019a,giacomini2019b,delahamette2020} shows that establishing such an alignment is itself a physical task, typically requiring system exchange or interaction with a common environment. This exposes a conceptual tension: while the process-matrix framework aims to describe correlations without a predefined causal order, its standard representation tacitly relies on reference structure whose operational status is left unaccounted for.

Recent work by Parker and Costa~\cite{parker2022} addresses one aspect of background independence by imposing invariance under global permutations of laboratory labels, ensuring that no laboratory is privileged by its identity alone. However, this symmetry acts only on labels and does not address the kinematical issue associated with local reference-frame alignment implicit in the CJ representation.

In this work, we analyze how this background structure can be removed operationally in a DI setting. We proceed in two logically coupled steps. First, we impose \emph{local-frame covariance} by requiring that operational predictions be invariant under independent local actions of the relevant physical symmetry group $G$ (e.g.\ $SU(2)$ for spin or $U(1)$ for phase). This is implemented via a $G\times G$ twirl over frame orientations, which removes any shared reference-frame background and induces a representation-theoretic decomposition into symmetry sectors and symmetry-invariant multiplicity subsystems carrying relational degrees of freedom. Second, we address the operational consequences of this decomposition at a \emph{fixed} device-independent operational interface specified by a causal game. Because symmetry-induced labels are not among the inputs or outputs specified by the causal game, they are treated as interface-excluded degrees of freedom, and causal signatures are analyzed at the corresponding coarse-grained operational interface.

At this interface, excluded degrees of freedom may still influence the reported statistics. We therefore introduce an \emph{interface-admissibility} condition, formalized in Sec.~\ref{sec:op-inter}, which distinguishes unassisted device-independent certification from scenarios involving unrecorded, setting-dependent control mediated by interface-excluded variables. A device-independent certification is called \emph{background-free} if it is obtained from a locally covariant implementation and satisfies this interface-admissibility requirement. This definition leads to the following operational question: which structural features must a locally covariant implementation possess in order to support a background-free device-independent violation of a causal inequality at a fixed coarse-grained interface?

We show that the answer hinges on the structure of the symmetry-invariant relational degrees of freedom. In the multiplicity-free regime, the covariant algebra is commutative and supports only static classical sector weights; with classical–classical (CC) multiplicity, internal classical memory or signalling may exist at the level of the physical implementation but becomes operationally inaccessible under background-free coarse-graining. In both cases, no background-free device-independent violation is possible: any apparent violation must rely on background assistance, such as symmetry breaking or unrecorded setting-dependent control. By contrast, non-CC multiplicity provides the only symmetry-invariant setting in which background-free violations can arise at the fixed coarse-grained interface, and we identify a sufficient structural condition based on embeddability of an effective input--output process-matrix structure into a non-CC multiplicity block. These results clarify which physical resources can enable device-independent causal signatures once reference-frame background and interface-level hidden control are excluded. 

The remainder of this paper is structured as follows. Section~\ref{sec:pm} reviews the process-matrix framework. Section~\ref{sec:lfc} introduces local-frame covariance and the induced invariant block structure. Section~\ref{sec:op-inter} formalizes operational interfaces and interface-admissible coarse-grained reporting. Our main no-go and structural results on necessary and sufficient conditions for causal-inequality violations are presented in Sec.~\ref{sec:bgfree}. We illustrate the framework in Sec.~\ref{sec:su2} and conclude with a discussion of implications and limitations in Sec.~\ref{sec:discussion}.

\section{Process-matrix framework}
\label{sec:pm}

Local laboratories Alice ($A$) and Bob ($B$) are modeled as quantum systems with input systems $A_I,B_I$ and output systems $A_O,B_O$, associated with finite-dimensional Hilbert spaces $\mathcal H_{A_I},\mathcal H_{A_O},\mathcal H_{B_I},\mathcal H_{B_O}$. Local operations are described by quantum instruments, i.e.\ collections of completely positive (CP) maps $\{M_{a|x}^A\}$ and $\{M_{b|y}^B\}$ whose sums are trace-preserving. Using the CJ representation, such maps are identified with positive operators acting on $\mathcal H_{A_I}\otimes\mathcal H_{A_O}$ and $\mathcal H_{B_I}\otimes\mathcal H_{B_O}$, respectively.

Given an operator $W\in\mathcal L(\mathcal H)$ acting on the composite Hilbert space
\[
\mathcal H
=
\mathcal H_{A_I} \otimes \mathcal H_{A_O}
\otimes
\mathcal H_{B_I} \otimes \mathcal H_{B_O},
\]
the generalized Born rule assigns joint probabilities
\begin{equation}
 p(a,b|x,y) = \Tr\!\left[W\,(M_{a|x}^A \otimes M_{b|y}^B)\right].
\label{eq:born-standard}
\end{equation}
Requiring well-defined probabilities for all local instruments imposes linear positivity, normalization, and causal-consistency constraints on $W$~\cite{oreshkov2012,araujo2015}.

\medskip

\emph{Linear constraints.}
It is convenient to express these constraints using the standard \emph{trace-and-replace} maps~\cite{araujo2015}. For any subsystem $X$, define
\begin{equation}
{}_X(W)
:=
\frac{\id_X}{d_X}\otimes \Tr_X(W).
\end{equation}
For composite subsystems $XY$, we define ${}_{XY}:={}_X\circ {}_Y$. A bipartite operator $W\in\mathcal L(\mathcal H)$ satisfies the process-matrix constraints if and only if:
\begin{enumerate}
\item $W\succeq 0$,
\item $\Tr(W)=d_{A_O}d_{B_O}$,
\item $W$ obeys the linear causal-consistency conditions
\begin{align}
W &= {}_{A_O}(W)+{}_{B_O}(W)-{}_{A_O B_O}(W), \label{eq:pm-out}\\
{}_{B_I B_O}(W) &= {}_{A_O B_I B_O}(W), \label{eq:pm-nsig-B}\\
{}_{A_I A_O}(W) &= {}_{A_I A_O B_O}(W). \label{eq:pm-nsig-A}
\end{align}
\end{enumerate}
The linear constraints~\eqref{eq:pm-out}--\eqref{eq:pm-nsig-A} define a subspace of $\mathcal{L}(\mathcal{H})$. The orthogonal projector $L_{\mathrm{PM}}$ onto this subspace is obtained by composing the commuting trace-and-replace projectors~\cite{araujo2015}.

\begin{definition}[Process matrix]
A \emph{(bipartite) process matrix} is an operator
$W\in\mathcal L(\mathcal H)$ that is positive semidefinite, correctly normalized, and satisfies
\begin{equation}
L_{\mathrm{PM}}(W)=W .
\end{equation}
\end{definition}

We now recall how causal structure is characterized within the set of admissible process matrices.

\begin{definition}[Causal order and causal separability]
A process matrix $W$ has a \emph{definite causal order} $A\prec B$ if Bob cannot signal to Alice, and analogously $B\prec A$ if Alice cannot signal to Bob. A process is \emph{causally separable} if it can be written as a convex mixture
\begin{equation}
 W_{\mathrm{sep}} = p\,W^{A\prec B} + (1-p)\,W^{B\prec A},
 \qquad p \in [0,1].
\end{equation}
\end{definition}

\medskip

\emph{Representation and implicit background structure.}
The standard process-matrix formulation~\cite{oreshkov2012} represents local operations via the CJ isomorphism, which requires fixing a correspondence between linear maps and operators on tensor-product Hilbert spaces. This correspondence is typically established by choosing a canonical maximally entangled state, such as $|\phi^+\rangle=\sum_i |i\rangle\otimes|i\rangle$, to identify input and output operator spaces. While this choice appears purely representational, adopting the same CJ convention across all laboratories implicitly selects a common description of local input--output spaces. In concrete physical implementations---for instance, when encoding quantum information in spin or phase degrees of freedom---this shared convention amounts to an effective alignment of local reference frames.

As noted in the Introduction, whether such an alignment exists physically and how it was established is left unspecified in standard DI analyses, yet establishing frame alignment is itself a resource-consuming physical process. This reveals residual background structure: the standard CJ representation tacitly assumes reference alignment whose physical origin is not accounted for. In Section~\ref{sec:lfc}, we address this by imposing \emph{local-frame covariance}---invariance under independent physical symmetry transformations acting on each laboratory---which removes shared reference-frame background by construction and exposes the algebraic and operational structure of genuinely frame-independent causal correlations.

\section{Local-Frame Covariance}
\label{sec:lfc}

We impose \emph{local-frame covariance}, since in the absence of a shared reference frame between independent laboratories, only relational, frame-independent correlations are operationally meaningful. Enforcing this principle has two distinct consequences. First, at the algebraic level, it constrains admissible process matrices to those invariant under independent local symmetry actions, inducing a representation-theoretic decomposition into symmetry sectors and multiplicity subsystems. Second, at the operational level, this decomposition introduces invariant labels that are not canonically part of the observed data, making DI causal claims sensitive to how the operational interface is specified. We analyze these consequences in sequence and show how they constrain DI causal-inequality violations across three structurally distinct regimes.

\subsection{Local-frame action and covariant processes}

Each laboratory carries a physical reference frame modeled by a compact Lie group $G$ determined by the symmetries of the underlying physical system. A frame transformation $g\in G$ acts on states of a system $X$ as
\begin{equation}
\rho \;\mapsto\; u_X(g)\,\rho\,u_X(g)^\dagger ,
\end{equation}
where $u_X(g)$ is a unitary representation of $G$ on $\mathcal H_X$. Within a given laboratory, the same local frame acts jointly on input and output systems.

We adopt a CJ convention in which the input space carries the contragredient (dual) representation relative to the output space. Accordingly, independent local frame transformations $(g_A,g_B)\in G\times G$ act on a bipartite process matrix $W$ by conjugation with
\begin{equation}
\label{eq:local-frame-action}
U(g_A,g_B)
=
u_{A_I}(g_A)^* \otimes u_{A_O}(g_A)
\otimes
u_{B_I}(g_B)^* \otimes u_{B_O}(g_B),
\end{equation}
i.e., $W \mapsto U(g_A,g_B)\,W\,U(g_A,g_B)^\dagger$. The absence of a shared reference frame is enforced by averaging over all local frame orientations (twirling):
\begin{equation}
\label{eq:twirl-def}
\mathcal T_{G\times G}(W)
:=
\int_{G\times G} dg_A\,dg_B\;
U(g_A,g_B)\,W\,U(g_A,g_B)^\dagger .
\end{equation}
This map is completely positive, trace preserving, and idempotent. It projects onto the subspace of frame-averaged processes, corresponding to operational scenarios where no shared reference frame is established between laboratories.

\begin{definition}[Frame-covariant process]
A bipartite process matrix $W$ is \emph{locally frame covariant} if $\mathcal T_{G\times G}(W)=W$. We denote the set of such processes by $\Wcov$.
\end{definition}

The process-matrix causal-consistency constraints are preserved by the symmetry operation.

\begin{lemma}[Covariance of causal-consistency constraints]
\label{lem-sr}
For all $(g_A,g_B)\in G\times G$,
\begin{equation}
\begin{aligned}
L_{\mathrm{PM}}\!\big(U(g_A,g_B)\,W\,U(g_A,g_B)^\dagger\big)
\\
=\;
U(g_A,g_B)\,L_{\mathrm{PM}}(W)\,U(g_A,g_B)^\dagger .
\end{aligned}
\end{equation}
Consequently, $\mathcal T_{G\times G}(W)$ is a valid process whenever $W$ is valid, so $\Wcov$ forms a covariant subspace of the space of valid process matrices.
\end{lemma}

\begin{proof}
Each trace-and-replace map ${}_X(\cdot)$ commutes with conjugation by unitaries that act as tensor products across laboratory factors: the partial trace eliminates the unitary on $X$, and replacement by $\id_X$ is invariant under conjugation. Since $L_{\mathrm{PM}}$ is a linear combination of such maps, it commutes with the $G\times G$ action, and hence with $\mathcal T_{G\times G}$.
\end{proof}

\subsection{Twirled structure: symmetry sectors and multiplicity subsystems}

The action of $G$ on each laboratory CJ space $\mathcal H_{X_I}\otimes\mathcal H_{X_O}$ defines a finite-dimensional unitary representation, which admits an isotypic decomposition into irreducible representations (irreps):
\begin{align}
\mathcal H_{A_I}\otimes\mathcal H_{A_O}
&\cong
\bigoplus_{j_A}
\mathcal R^A_{j_A}\otimes \mathcal M^A_{j_A},
\nonumber\\
\mathcal H_{B_I}\otimes\mathcal H_{B_O}
&\cong
\bigoplus_{j_B}
\mathcal R^B_{j_B}\otimes \mathcal M^B_{j_B}.
\label{eq:lab-decomp}
\end{align}
Here $\mathcal R^{A,B}_{j}$ carry irreps of $G$ and $\mathcal M^{A,B}_{j}$ are multiplicity spaces on which $G$ acts trivially. We define the joint multiplicity space $\mathcal M_{j_A j_B}:=\mathcal M^A_{j_A}\otimes\mathcal M^B_{j_B}$.

\begin{theorem}[Invariant block form]
\label{thm-ibf}
An operator $W$ is locally frame covariant if and only if it can be written as
\begin{equation}
W
=
\bigoplus_{j_A,j_B}
\left(
\frac{\id_{\mathcal R^A_{j_A}}}{d_{j_A}}
\otimes
\frac{\id_{\mathcal R^B_{j_B}}}{d_{j_B}}
\right)
\otimes
\widetilde W_{j_A j_B},
\label{eq:covariant-structure}
\end{equation}
where $d_{j_A}=\dim\mathcal R^A_{j_A}$, $d_{j_B}=\dim\mathcal R^B_{j_B}$, and each $\widetilde W_{j_A j_B}$ acts on $\mathcal M_{j_A j_B}$.
\end{theorem}

\begin{proof}
$W$ is locally frame covariant iff it commutes with $U(g_A,g_B)$ for all $(g_A,g_B)\in G\times G$. In the isotypic decomposition $\mathcal H_X\simeq\bigoplus_j(\mathcal R^X_j\otimes\mathcal M^X_j)$, Schur's lemma implies that the commutant of the $G$ action consists exactly of operators that are block-diagonal in $j$, proportional to the identity on each $\mathcal R^X_j$, and arbitrary on $\mathcal M^X_j$. Since the global action factorizes across laboratories, invariance under $G\times G$ yields the stated block form. The factors $\id/d$ are a convenient normalization convention.
\end{proof}

Theorem~\ref{thm-ibf} identifies the twirl-invariant operator algebra with the direct sum of multiplicity algebras,
\[
\Wcov \;\cong\; \bigoplus_{j_A,j_B} \mathcal L(\mathcal M_{j_A j_B}),
\]
where $W\in\Wcov$ is uniquely specified by the collection $(\widetilde W_{j_A j_B})_{j_A,j_B}$.

Define the canonical blockwise embedding $\Gamma:\bigoplus_{j_A,j_B}\mathcal L(\mathcal M_{j_A j_B})\to\mathcal L(\mathcal H)$ by
\begin{equation}
\label{eq:Gamma-def}
\Gamma(\widetilde W)
:=
\bigoplus_{j_A,j_B}
\left(
\frac{\id_{\mathcal R^A_{j_A}}}{d_{j_A}}
\otimes
\frac{\id_{\mathcal R^B_{j_B}}}{d_{j_B}}
\right)
\otimes
\bigl(\widetilde W\bigr)_{j_A j_B}.
\end{equation}
By Lemma~\ref{lem-sr}, $L_{\mathrm{PM}}$ preserves $\Wcov$, and thus induces a unique linear map $\widetilde L_{\mathrm{PM}}$ on the multiplicity algebra via $\Gamma\!\bigl(\widetilde L_{\mathrm{PM}}(\widetilde W)\bigr)=L_{\mathrm{PM}}\!\bigl(\Gamma(\widetilde W)\bigr)$. Admissibility of covariant processes therefore reduces to the fixed-point condition $\widetilde L_{\mathrm{PM}}(\widetilde W)=\widetilde W$ (together with positivity and normalization) on the multiplicity algebra. Operationally, this ensures that one can impose the process-matrix causal-consistency constraints \emph{after} enforcing local-frame covariance, and analyze which covariant, symmetry-invariant degrees of freedom remain compatible with those constraints without reintroducing any symmetry-breaking structure.

\section{Operational Interfaces and Background-Free Coarse-Graining}
\label{sec:op-inter}

After enforcing local-frame covariance, the twirl induces additional invariant labels---the symmetry sectors $s:=(j_A,j_B)$---that are not canonically part of the observed dataset. DI causal signatures must therefore be defined relative to an explicit \emph{operational interface} specifying which variables are recorded and which are coarse-grained.

\paragraph{Resolved versus coarse-grained interfaces.}
Recording sector labels yields distributions such as $p(a,b,j_A,j_B|x,y)$ or $p(a,b|x,y,j_A,j_B)$, where conditioning on sectors is operationally legitimate because the labels are part of the recorded variables. Discarding sector labels yields the marginalized distribution
\[
p(a,b|x,y)=\sum_{j_A,j_B} p(a,b,j_A,j_B|x,y),
\]
in which sectors are excluded from the recorded data and any influence of sector information on observed statistics occurs only through the physical response of the devices.

Causal inequalities are inherently sensitive to the choice of operational interface. While sector-resolved statistics $p(a,b,s|x,y)$ fully specify how correlations depend on the symmetry labels, marginalization over $s$ discards information about how sector contributions depend on the settings. As a result, a violation observed at the sector-resolved level need not survive coarse-graining, and distinct physical mechanisms can lead to the same coarse-grained distribution. Any analysis based on coarse-grained data must therefore make explicit what is assumed about the behaviour of interface-excluded degrees of freedom.

\paragraph{Fixed DI task.}
For definiteness, we focus on DI causal games of the GYNI type~\cite{branciard2016}, whose operational interface consists solely of the classical settings and outcomes $(x,y,a,b)$ specified by the task. Symmetry-sector labels arising from local-frame covariance are treated as internal degrees of freedom of the physical implementation, not inputs or outputs of the game itself. Our results are therefore statements about which causal signatures remain certifiable at this fixed operational interface. While a covariant process may encode nonclassical causal structure at a sector-resolved level, such structure is operationally inaccessible once sector labels are excluded. Promoting these labels to accessible settings---for instance, by replacing $(x,y)$ with $(x,j_A; y,j_B)$---would enlarge the operational interface and thereby define a strictly different DI task and set of causal constraints. Such extended-interface scenarios are outside the scope of the present analysis.

\paragraph{Interface-admissible coarse-grained data.}
Fixing a coarse-grained operational interface does not, by itself, determine how interface-excluded degrees of freedom may influence the reported statistics. The following definition specifies the minimal conditions under which coarse-grained distribution $p(a,b|x,y)$ admits a DI interpretation.

\begin{definition}[Interface-admissible coarse-grained DI statistics]
\label{def:interface_admissible}
Consider a coarse-grained operational interface in which the symmetry-induced label $s$ is excluded from the recorded data and only the distribution $p(a,b|x,y)$ is reported. Let $p(a,b,s|x,y)$ denote the joint distribution induced by the physical implementation prior to coarse-graining, so that
\[
p(a,b|x,y)=\sum_s p(a,b,s|x,y).
\]

The reported behaviour $p(a,b|x,y)$ is \emph{interface-admissible} if the following conditions hold.
\begin{enumerate}
\item[\textnormal{(NC)}] \textbf{No hidden conditioning or renormalization.}
All experimental runs are included in the analyzed dataset. Any loss, failure, or abort event is treated as part of the recorded outcomes (e.g.\ via an explicit outcome $\bot$), so that no renormalization depending on $s$ is performed.

\item[\textnormal{(IC)}] \textbf{Interface consistency.}
The marginal contribution of any interface-excluded label must be independent of the settings. Equivalently, there exists a distribution $p(s)$ such that
\begin{equation}
\label{eq:IC_sectorweights}
\sum_{a,b} p(a,b,s|x,y) = p(s)
\qquad \forall x,y .
\end{equation}
\end{enumerate}
\end{definition}

\emph{Operational interpretation.}
Interface-admissibility is an identifiability requirement that specifies which 
coarse-grained datasets admit a DI interpretation without relying 
on unrecorded interface-level control. It imposes no constraints on the underlying 
physics: local instruments may be non-covariant, and physical noise may freely mix 
symmetry sectors. Rather, it acts as a logical prerequisite for certification that 
cannot be verified from coarse-grained data alone but must be justified through 
experimental design, calibration, or trusted symmetrization procedures. Behaviours 
violating either (NC) or (IC) are called \emph{interface-assisted}.

Condition (NC) is standard in DI reasoning: it excludes hidden 
postselection or renormalization through unrecorded loss, heralding, or abort 
events, ensuring all experimental runs contribute to the reported statistics.

Condition (IC) addresses a distinct issue arising at coarse-grained interfaces. 
At the physical level, different settings $(x,y)$ correspond to different CPTP 
maps and may legitimately affect the sector label $s$ through sector-dependent 
dynamics, loss, or mixing. However, since the declared DI interface consists 
solely of $(x,y,a,b)$, any setting dependence of the marginal weights $p(s|x,y)=\sum_{a,b}p(a,b,s|x,y)$ remains unobservable in the reported data $p(a,b|x,y)$. This creates an identifiability failure: whenever $p(s|x,y)$ depends on $(x,y)$, the same 
coarse-grained behaviour admits an operationally indistinguishable realization 
via hidden routing. An adversary could sample $s\sim p(s|x,y)$ and generate 
outputs according to $p(a,b|x,y,s)=p(a,b,s|x,y)/p(s|x,y)$, and no test on 
$(x,y,a,b)$ alone could detect this intervention. Condition (IC) rules out this 
ambiguity by requiring $p(s|x,y)=p(s)$.

This is directly analogous to measurement independence in Bell scenarios. There, 
the requirement $p(\lambda|x,y)=p(\lambda)$ prevents hidden signalling via shared 
variable $\lambda$ from invalidating nonlocality certification---without it, the 
same observed correlations could arise from setting-dependent preparation of $\lambda$ 
rather than nonlocal influences. Here, (IC) plays the same role: it ensures that 
setting-dependent modulation of interface-excluded degrees of freedom cannot 
be confused with an unassisted causal signature. Crucially, $s$ is not an abstract 
hidden variable but a physically well-defined, symmetry-derived degree of freedom 
that may be observable in principle. Once excluded from the operational interface, 
however, its physical origin becomes irrelevant for DI certification: 
any setting dependence is operationally indistinguishable from hidden control.

Throughout this work, a DI certification is called \emph{background-free} if and only if it arises from a locally covariant physical implementation and the reported data are interface-admissible. This joint requirement excludes both symmetry-breaking reference-frame resources and hidden interface-level control. While interface-assisted or symmetry-breaking implementations may be physically valid and may generate causal-inequality violations, such violations are not operationally attributable to symmetry-invariant relational structure at the level of the reported data.

\section{Background-Free Causal Signatures}
\label{sec:bgfree}

We now analyze which DI causal signatures can survive local-frame covariance when evaluated at a fixed coarse-grained operational interface under the interface-admissibility conditions of Definition~\ref{def:interface_admissible}.

\subsection{Regime I: multiplicity-free covariant data}

We begin with the simplest covariant regime, in which the symmetry-induced decomposition contains no nontrivial multiplicity subsystems.

\begin{corollary}[Multiplicity-free structure]
\label{cor-mfc}
Assume that the multiplicity spaces are trivial, $\dim \mathcal M^A_{j_A} = 1$ and $\dim \mathcal M^B_{j_B} = 1$ for all $j_A,j_B$. Then every frame-independent process $W\in\Wcov$ is of the form
\begin{equation}
W
=
\bigoplus_{j_A,j_B}
\lambda_{j_A j_B}
\left(
\frac{\id_{\mathcal R^A_{j_A}}}{d_{j_A}}
\otimes
\frac{\id_{\mathcal R^B_{j_B}}}{d_{j_B}}
\right),
\qquad
\lambda_{j_A j_B}\ge 0 ,
\label{eq:sector-classical}
\end{equation}
and the invariant algebra $\Wcov$ is commutative.
\end{corollary}

\begin{proof}
When $\dim \mathcal M_{j_A j_B}=1$, each $\widetilde W_{j_A j_B}$ in~\eqref{eq:covariant-structure} reduces to a nonnegative scalar $\lambda_{j_A j_B}$, yielding~\eqref{eq:sector-classical}. The resulting invariant algebra is a direct sum of one-dimensional blocks and is therefore commutative.
\end{proof}

\begin{proposition}[Multiplicity-free covariant data admit no background-free DI causal violation]
\label{prop:mf_bgfree_noCI}
In the multiplicity-free (MF) regime, consider the coarse-grained interface $p(a,b|x,y)$ and assume interface-admissible coarse-grained data in the sense of Definition~\ref{def:interface_admissible}. Then the reported behaviour admits a normalized convex-mixture decomposition of product response functions at the fixed interface; in particular, no background-free DI violation of any bipartite causal inequality is possible at this interface.
\end{proposition}

\begin{proof}
By Corollary~\ref{cor-mfc}, in the multiplicity-free regime any covariant process $W\in\Wcov$ decomposes as a direct sum over symmetry sectors $s=(j_A,j_B)$, and on each sector acts as a scalar multiple of the identity on $\mathcal R^A_{j_A}\otimes\mathcal R^B_{j_B}$. For each $s=(j_A,j_B)$ define
\[
W_s:=\lambda_s\left(\frac{\id_{\mathcal R^A_{j_A}}}{d_{j_A}}\otimes\frac{\id_{\mathcal R^B_{j_B}}}{d_{j_B}}\right),
\]
so that $W=\bigoplus_s W_s$. This induces a well-defined pre-coarse-grained joint distribution
\[
p(a,b,s|x,y)
:=\Tr\!\left[W_s\,(M^A_{a|x}\otimes M^B_{b|y})\right],
\]
\[
p(a,b|x,y)
=\sum_s p(a,b,s|x,y).
\]
matching the role of $p(a,b,s|x,y)$ in Definition~\ref{def:interface_admissible}.

Let $\Pi^A_{j_A}$ and $\Pi^B_{j_B}$ denote the projectors
onto the corresponding isotypic components of the local Choi spaces. For
arbitrary instruments $\{M^A_{a|x}\}_a$ and $\{M^B_{b|y}\}_b$, the observed
probabilities read
\[
\begin{aligned}
p(a,b|x,y)
&= \Tr\!\left[W\,(M^A_{a|x}\otimes M^B_{b|y})\right] \\
&= \sum_{s} \lambda_s\,\alpha_s(a|x)\,\beta_s(b|y).
\end{aligned}
\]
where $\lambda_s:=\lambda_{j_A j_B}\ge0$ and
\[
\begin{aligned}
\alpha_s(a|x)
&:= \Tr\!\left[M^A_{a|x}\Big(\frac{\Pi^A_{j_A}}{d_{j_A}}\Big)\right], \\
\beta_s(b|y)
&:= \Tr\!\left[M^B_{b|y}\Big(\frac{\Pi^B_{j_B}}{d_{j_B}}\Big)\right].
\end{aligned}
\]

Define the sector totals
\[
\begin{aligned}
A_s(x)
&:= \sum_a \alpha_s(a|x)
= \Tr\!\left[M^A_x\Big(\frac{\Pi^A_{j_A}}{d_{j_A}}\Big)\right], \\
B_s(y)
&:= \sum_b \beta_s(b|y)
= \Tr\!\left[M^B_y\Big(\frac{\Pi^B_{j_B}}{d_{j_B}}\Big)\right].
\end{aligned}
\]
with $M^A_x:=\sum_a M^A_{a|x}$ and $M^B_y:=\sum_b M^B_{b|y}$. Summing over outcomes factorizes because $W_s$ is a product operator on the two laboratories, hence the sector weights satisfy
\begin{equation}
\label{eq:psxy_factor}
p(s|x,y)=\sum_{a,b} p(a,b,s|x,y)=\lambda_s\,A_s(x)\,B_s(y).
\end{equation}

Condition (IC) requires $p(s|x,y)=p(s)$ for all $(x,y)$, which enforces that the
product $\lambda_s A_s(x)B_s(y)$ is independent of the settings. Sectors for
which $A_s(x)B_s(y)=0$ for all $(x,y)$ contribute zero probability and can be
discarded. For the remaining sectors, define normalized local responses
\[
p_A(a|x,s):=\frac{\alpha_s(a|x)}{A_s(x)}, \qquad
p_B(b|y,s):=\frac{\beta_s(b|y)}{B_s(y)},
\]
and the setting-independent sector weight
\[
q(s):=p(s)=\lambda_s A_s(x)B_s(y).
\]

By condition (NC), all experimental runs are included in the reported data and
no setting-dependent renormalization is performed. Consequently, the
coarse-grained distribution is normalized for all $(x,y)$:
\[
1=\sum_{a,b} p(a,b|x,y)
= \sum_s \lambda_s A_s B_s
= \sum_s q(s),
\]
so $\tilde q$ is a bona fide probability distribution. Substituting the above
definitions yields
\[
p(a,b|x,y)=\sum_s q(s)\,p_A(a|x,s)\,p_B(b|y,s).
\]

This is a convex mixture of product response functions at the fixed interface,
i.e.\ a shared-randomness model over the excluded label $s$. In particular,
the reported behaviour cannot violate any bipartite causal inequality.
\end{proof}

\subsection{Regime II: classical-classical multiplicity}

We now allow nontrivial multiplicity spaces and consider the case where the multiplicity blocks of the \emph{process instance} are classical--classical (CC). Concretely, for all $j_A,j_B$ there exist local orthonormal bases $\{|k\rangle\}$ of $\mathcal M^A_{j_A}$ and $\{|\ell\rangle\}$ of $\mathcal M^B_{j_B}$ such that
\begin{equation}
\widetilde W_{j_Aj_B}
=
\sum_{k,\ell} \mu^{j_Aj_B}_{k\ell}\,
|k\rangle\!\langle k|_{\mathcal M^A_{j_A}}
\otimes
|\ell\rangle\!\langle \ell|_{\mathcal M^B_{j_B}},
\qquad \mu^{j_Aj_B}_{k\ell}\ge0.
\label{eq:cc_mult}
\end{equation}

At the level of observable probabilities, the CC case reduces to the MF case with a refined classical hidden label, corresponding to the joint diagonalization of the multiplicity blocks.

\begin{proposition}[CC multiplicity admits no background-free DI causal-inequality violation]
\label{prop:cc_bgfree_noCI}
Assume the coarse-grained interface and interface-admissible statistics as in Definition~\ref{def:interface_admissible}. If, in addition, all multiplicity blocks are classical--classical as in~\eqref{eq:cc_mult}, then the reported behaviour admits a normalized convex-mixture decomposition of product response functions at the fixed interface, and therefore no background-free DI violation of any bipartite causal inequality is possible at this interface.
\end{proposition}

\begin{proof}
By Theorem~\ref{thm-ibf}, any locally frame-independent $W$ is block-diagonal in $(j_A,j_B)$ with blocks $\widetilde W_{j_Aj_B}$ acting on $\mathcal M^A_{j_A}\otimes \mathcal M^B_{j_B}$. Under the CC assumption~\eqref{eq:cc_mult}, each $\widetilde W_{j_Aj_B}$ is diagonal in a local product basis, so the full process admits a decomposition as a direct sum over a global
classical label $r:=(j_A,j_B,k,\ell)$, with no coherence between distinct values of $r$.

Writing $W=\bigoplus_r W_r$ for the corresponding positive diagonal blocks, we define the pre-coarse-grained distribution $p(a,b,r|x,y):=\Tr[W_r\,(M^A_{a|x}\otimes M^B_{b|y})]$, with $p(a,b|x,y)=\sum_r p(a,b,r|x,y)$, so that $r$ plays the role of an interface-excluded label in the sense of Definition~\ref{def:interface_admissible}.

For arbitrary instruments, the Born rule yields a decomposition $p(a,b|x,y)=\sum_r \alpha_r(a|x)\,\beta_r(b|y)$, where $\alpha_r(a|x),\beta_r(b|y)\ge0$ are generally subnormalized. Define the marginal sector weights $p(r|x,y):=\sum_{a,b} p(a,b,r|x,y)$. Applying Definition~\ref{def:interface_admissible}, condition (IC) enforces $p(r|x,y)=p(r)$ for all $(x,y)$. Discarding sectors with $p(r)=0$, we define normalized local response functions $p_A(a|x,r):=\alpha_r(a|x)/\sum_a \alpha_r(a|x)$ and $p_B(b|y,r):=\beta_r(b|y)/\sum_b \beta_r(b|y)$, and set $q(r):=p(r)$. Condition (NC) guarantees that no setting-dependent renormalization occurs, so $\sum_r q(r)=1$.

Substituting, we obtain $p(a,b|x,y)=\sum_r q(r)\,p_A(a|x,r)\,p_B(b|y,r)$, which is a normalized convex mixture of product response functions at the fixed operational interface,
i.e.\ a shared-randomness model over the excluded label $r$. Hence $p(a,b|x,y)$ cannot violate any bipartite causal inequality.
\end{proof}

\subsection{Regime III: quantum multiplicity}

We now turn to the only remaining structural regime compatible with a genuinely
background-free DI violation of a causal inequality.
Suppose that for at least one pair $(j_A,j_B)$ the invariant block
$\widetilde W_{j_A j_B}$ acting on
$\mathcal M^A_{j_A}\otimes\mathcal M^B_{j_B}$ is \emph{not}
classical--classical (non-CC), i.e.\ it cannot be diagonalized in a product basis
by local unitary changes on $\mathcal M^A_{j_A}$ and $\mathcal M^B_{j_B}$.

In this case, the covariant process instance contains symmetry-invariant
quantum relational degrees of freedom that cannot be reduced
to a joint classical register within each symmetry sector.

\begin{corollary}[Non-CC multiplicity is necessary for background-free DI violations]
\label{cor:noncc_necessary}
Under the coarse-grained operational interface and interface-admissible statistics of
Definition~\ref{def:interface_admissible}, any DI violation of a bipartite
causal inequality requires that the employed covariant process has
non-classical--classical (non-CC) multiplicity in at least one invariant block.
\end{corollary}

\begin{proof}
If the invariant structure is multiplicity free, the conclusion follows from
Proposition~\ref{prop:mf_bgfree_noCI}.
If multiplicity is present but all invariant blocks are classical--classical, the
conclusion follows from Proposition~\ref{prop:cc_bgfree_noCI}.
Therefore, any background-free DI causal-inequality violation must involve at
least one non-CC multiplicity block.
\end{proof}

The absence of causal-inequality violations in the MF and CC regimes should not be
interpreted as an absence of physical influence or internal dynamics. In both
cases, signalling or memory may exist among interface-excluded degrees of freedom,
but under background-free coarse-graining such influences are not operationally
accessible at the fixed DI interface. Non-CC multiplicity is therefore required not to enable signalling per se. Rather, it provides genuinely quantum relational structure capable of producing DI causal-inequality violations without hidden control or symmetry-breaking background assistance.

Corollary~\ref{cor:noncc_necessary} is a necessity statement only. It identifies
non-CC multiplicity as the unique structural regime in which genuinely
background-free DI violations can occur, but it does not assert that every such
process instance yields a violation. Whether non-CC invariant structure yields an observable background-free DI causal-inequality violation depends on whether the corresponding multiplicity block admits a solution of the reduced process-matrix fixed-point constraints that remains compatible with interface-admissible coarse-grained reporting. We now show that this regime is not ruled out by providing a sufficient structural condition under which a non-CC multiplicity block can support a background-free DI causal-inequality violation.

\subsection{A sufficient structural condition: IO-embeddability into multiplicity}

We now identify a sufficient structural condition under which a non-CC multiplicity block can support a background-free DI causal-inequality violation. Specifically, if such a block admits an effective input--output (IO) structure compatible with the reduced process-matrix constraints, then causally nonseparable processes can be realized entirely within symmetry-invariant relational degrees of freedom.

\begin{proposition}[Existence of background-free DI violations via IO-embeddability]
\label{prop:existence_embedding}
Suppose there exists an invariant block $(j_A,j_B)$ and an effective bipartite process matrix $W_{\mathrm{eff}}$ acting on finite-dimensional input and output spaces $\mathcal H_{A_I}^{\mathrm{eff}},\mathcal H_{A_O}^{\mathrm{eff}},\mathcal H_{B_I}^{\mathrm{eff}},\mathcal H_{B_O}^{\mathrm{eff}}$ such that:
\begin{enumerate}
\item[(i)] There exist isometries $V_A:\mathcal H_{A_I}^{\mathrm{eff}}\otimes\mathcal H_{A_O}^{\mathrm{eff}}\hookrightarrow \mathcal M^A_{j_A}$ and $V_B:\mathcal H_{B_I}^{\mathrm{eff}}\otimes\mathcal H_{B_O}^{\mathrm{eff}}\hookrightarrow \mathcal M^B_{j_B}$ for which the embedded operator $\widetilde W_{j_A j_B}:=(V_A\otimes V_B)\,W_{\mathrm{eff}}\,(V_A\otimes V_B)^\dagger$ is compatible with the reduced process-matrix constraints on the multiplicity algebra, i.e.\ satisfies $\widetilde L_{\mathrm{PM}}(\widetilde W_{j_A j_B})=\widetilde W_{j_A j_B}$.

\item[(ii)] There exist effective instruments $\{M^{A,\mathrm{eff}}_{a|x}\},\{M^{B,\mathrm{eff}}_{b|y}\}$ that witness a bipartite causal-inequality violation for $W_{\mathrm{eff}}$, and whose lifted counterparts can be implemented on the full multiplicity spaces without postselection or setting-dependent renormalization, so that Definition~\ref{def:interface_admissible} is satisfied.
\end{enumerate}
Then there exists a locally frame-covariant process matrix whose coarse-grained behaviour $p(a,b|x,y)$ violates a bipartite causal inequality in a background-free DI manner.
\end{proposition}

\begin{proof}
Let $W_{\mathrm{eff}}$ be a bipartite process matrix violating a causal inequality in a DI causal game. By assumption (i), the operator $\widetilde W_{j_A j_B}$ defines an admissible invariant block on $\mathcal M^A_{j_A}\otimes\mathcal M^B_{j_B}$. Define a covariant process matrix by embedding this block and setting all other invariant blocks to zero, namely
\[
W=\Gamma(\widetilde W)=\bigoplus_{j'_A,j'_B}\left(\frac{\id_{\mathcal R^A_{j'_A}}}{d_{j'_A}}\otimes\frac{\id_{\mathcal R^B_{j'_B}}}{d_{j'_B}}\otimes\widetilde W_{j'_A j'_B}\right),
\]
with $\widetilde W_{j'_A j'_B}=\widetilde W_{j_A j_B}$ if $(j'_A,j'_B)=(j_A,j_B)$ and zero otherwise. By construction, $W$ is locally frame covariant and satisfies the full process-matrix constraints.

By assumption (ii), the lifted instruments generate background-free coarse-grained statistics at the fixed operational interface, and the resulting probabilities coincide with those of the effective process, namely $p(a,b|x,y)=\Tr\!\left[W_{\mathrm{eff}}\,(M^{A,\mathrm{eff}}_{a|x}\otimes M^{B,\mathrm{eff}}_{b|y})\right]$. Since $W_{\mathrm{eff}}$ violates a bipartite causal inequality, the induced coarse-grained behaviour does so as well.
\end{proof}

\emph{On the IO-embeddability condition.}
In the construction above, all nonclassical causal structure is encoded entirely within symmetry-invariant multiplicity degrees of freedom, without invoking hidden control, postselection, or symmetry-breaking resources. Proposition~\ref{prop:existence_embedding} thus establishes a \emph{possibility-in-principle}: enforcing local-frame covariance does not, by itself, preclude genuinely background-free DI violations of causal inequalities. Together with the necessity statement of Corollary~\ref{cor:noncc_necessary}, this singles out non-CC multiplicity as the unique symmetry-invariant resource capable of supporting such violations.

The role of IO-embeddability is to characterize when this invariant relational structure is not merely present, but sufficiently expressive to host an effective input--output process compatible with the reduced process-matrix constraints. This condition is inherently non-generic. Verifying embeddability for a given physical system requires analyzing how the reduced trace-and-replace constraints $\widetilde L_{\mathrm{PM}}$ act on the commutant algebra, and whether a CJ-like subalgebra closed under causal-consistency constraints exists within the multiplicity space. While larger or more structured multiplicity spaces make embeddability more plausible, constructing explicit examples or deriving general representation-theoretic criteria is left for future work.

\section{Illustration: $SU(2)$-Covariant Spin Laboratories}
\label{sec:su2}

We illustrate the framework with laboratories whose local degrees of freedom are encoded into $N$ spin-$\tfrac12$ particles, in the absence of a shared spatial reference frame. Local frame transformations are modeled by independent $SU(2)_A$ and $SU(2)_B$ rotations acting on Alice’s and Bob’s laboratories, respectively, and local-frame covariance requires invariance under $SU(2)_A\times SU(2)_B$.

The laboratory Choi space $\mathcal H_{X_I}\otimes\mathcal H_{X_O}$ carries the representation $(u^{\otimes N})^*\otimes u^{\otimes N}$, which is unitarily equivalent to $u^{\otimes N}\otimes u^{\otimes N}$ for $SU(2)$. Enforcing local-frame covariance therefore induces a decomposition into total-spin sectors and associated multiplicity spaces, whose structure depends on $N$.

\emph{Single qubit per laboratory ($N=1$): multiplicity-free regime.}
For $N=1$, the local Choi-space representation decomposes as $\tfrac12\otimes\tfrac12\cong 0\oplus 1$, which is multiplicity free. By Corollary~\ref{cor-mfc}, any locally frame-independent process takes the form
\[
W=\sum_{j_A,j_B\in\{0,1\}} \lambda_{j_A j_B}\left(\frac{\id_{\mathcal R^A_{j_A}}}{d_{j_A}}\otimes\frac{\id_{\mathcal R^B_{j_B}}}{d_{j_B}}\right),\qquad \lambda_{j_A j_B}\ge 0,
\]
with a commutative invariant algebra. The only $SU(2)_A\times SU(2)_B$-invariant structure is a classical distribution over sector labels $(j_A,j_B)$. By Proposition~\ref{prop:mf_bgfree_noCI}, under interface-admissible coarse-grained reporting (Definition~\ref{def:interface_admissible}) within a locally covariant implementation, the coarse-grained behaviour admits a convex-mixture decomposition of product response functions at the fixed interface and therefore cannot violate any bipartite causal inequality. 

Canonical constructions such as the Oreshkov--Costa--Brukner (OCB) process, when implemented with spins in a fixed basis, are not $SU(2)_A\times SU(2)_B$-covariant and therefore rely on symmetry-breaking resources. Alternatively, apparent nonclassical causal signatures may arise from unrecorded setting-dependent control at the operational interface. Both mechanisms fall outside the background-free covariant framework considered here.

\emph{Multiplicity and the CC vs non-CC distinction ($N\ge 2$).}
For $N\ge2$ qubits per laboratory, the local Choi-space representations acquire nontrivial multiplicity spaces. Already for $N=2$, the decomposition $(0\oplus1)\otimes(0\oplus1)=0^{\oplus2}\oplus1^{\oplus3}\oplus2$ contains repeated irreducible representations, so the twirl-invariant operator algebra becomes genuinely noncommutative,
\[
\Wcov \;\cong\; \bigoplus_{j_A,j_B} \mathcal L(\mathcal M_{j_A j_B}).
\]
This kinematical noncommutativity signals the availability of symmetry-invariant relational degrees of freedom, but it is not by itself sufficient to generate nonclassical background-free DI behaviour. What matters operationally is how a given covariant process instance populates the multiplicity blocks.

If the realized blocks are CC, i.e.\ jointly diagonalizable in local bases as in~\eqref{eq:cc_mult}, the invariant degrees of freedom reduce to a joint classical register. 
Under interface-admissible coarse-grained reporting within a locally covariant implementation, the resulting behaviour cannot violate any bipartite causal inequality, in accordance with Proposition~\ref{prop:cc_bgfree_noCI}. In this case, the presence of multiplicity does not translate into background-free DI nonclassical causal signatures.

The situation changes only when a process exploits non-CC structure within a multiplicity block. In that regime, genuinely quantum coherence can be encoded entirely within symmetry-invariant relational degrees of freedom, and the MF and CC no-go results no longer apply. Whether such structure yields an observable background-free DI causal-inequality violation depends on compatibility with the reduced process-matrix constraints $\widetilde L_{\mathrm{PM}}(\widetilde W)=\widetilde W$ and must be assessed case by case.

This $SU(2)$ example illustrates that symmetry-invariant multiplicity provides the
minimal structural setting in which background-free nonclassical causal structure
can arise once shared reference-frame resources are excluded. While local-frame
covariance removes any reliance on shared spatial orientation, it leaves open the
possibility of exploiting relational degrees of freedom encoded within
multiplicity subsystems. Whether such non-CC relational structure can be converted into an observable device-independent causal-inequality violation in concrete physical implementations remains a separate, constructive question not addressed in this work.

\section{Discussion}
\label{sec:discussion}

We have analyzed device-independent causal signatures under enforced local-frame
covariance, treating the absence of shared reference frames as an operational
constraint on admissible background structure. Imposing invariance under independent local symmetry actions $G\times G$ separates symmetry-breaking resources from symmetry-invariant
relational degrees of freedom. Combined with explicit specification of the operational interface and admissible coarse-grained reporting (Definition~\ref{def:interface_admissible}), this framework yields a sharp operational distinction between DI causal-inequality signatures that require background assistance and those that persist without it.

\emph{Multiplicity structure regimes.}
Our results identify three distinct regimes determined by the multiplicity
structure of the covariant process. In multiplicity-free (MF) settings,
local-frame covariance eliminates all nontrivial symmetry-invariant subsystems:
the covariant algebra reduces to an abelian structure supporting only
classical sector weights (Corollary~\ref{cor-mfc}). Operationally, the only
available invariant resource is shared classical randomness fixed at
preparation. Under interface-admissible coarse-grained reporting, these sector
weights are setting independent, and the resulting behaviour cannot violate any
bipartite causal inequality (Proposition~\ref{prop:mf_bgfree_noCI}). Any apparent
violation in this regime must therefore rely on background assistance, either
through symmetry-breaking resources that effectively establish a shared frame
or through hidden control mechanisms such as setting-dependent filtering over
interface-excluded degrees of freedom. Canonical constructions such as the OCB
process~\cite{oreshkov2012}, when implemented with qubit laboratories in a fixed
local basis, fall into this category.

When multiplicity subsystems are present, additional symmetry-invariant
structure becomes available. If the realized multiplicity blocks are
classical--classical (CC), this structure reduces to a joint classical register
shared between the parties. At the sector-resolved level, such a register can
support classical memory or signalling within the physical implementation.
However, under interface-admissible coarse-grained reporting, the sector weights
are required to be setting independent, so the influence of these internal
classical resources averages out in the reported statistics. As a result, CC
multiplicity still cannot support DI causal-inequality violations at the
coarse-grained interface (Proposition~\ref{prop:cc_bgfree_noCI}). Kinematical
noncommutativity of the invariant algebra is therefore not sufficient:
genuinely background-free DI violations require at least one
non-classical--classical (non-CC) multiplicity block in the realized covariant
process (Corollary~\ref{cor:noncc_necessary}). In this quantum regime, causal
influence can be encoded coherently within a fixed symmetry-invariant block,
without modulating sector weights and hence without violating the interface
constraint. Whether covariant processes with non-CC multiplicity generically
lead to background-free causal-inequality violations remains open. The present
work establishes the necessity of such structure and identifies one explicit
sufficient route---IO embeddability within multiplicity---for background-free DI
violations (Proposition~\ref{prop:existence_embedding}).

\emph{Covariance, interface admissibility, and dynamics.}
The multiplicity subsystems identified here arise purely kinematically as commutant degrees of freedom of the $G\times G$ action, and should not be interpreted as dynamical superselection sectors or decoherence-free subsystems. Enforcing local-frame covariance constrains the algebraic form of invariant operators but imposes no dynamical restrictions: local instruments need not be covariant, and physical noise may freely mix symmetry sectors or implement arbitrary CPTP dynamics within multiplicity spaces. The interface-admissible conditions of Definition~\ref{def:interface_admissible} likewise do not constrain the underlying dynamics. Instead, they act as certification-level identifiability conditions that delimit which reported
coarse-grained distributions admit an \emph{unassisted} device-independent interpretation at the fixed operational interface. Like measurement independence in Bell scenarios, these conditions cannot be certified from coarse-grained data alone and must instead be justified at the
level of the physical implementation. Our conclusions therefore concern which causal signatures can be attributed to symmetry-invariant relational structure without
hidden control, rather than how such structure is dynamically generated or protected.

\emph{Resource-theoretic classification.}
These results motivate a refined resource-theoretic interpretation that separates two logically distinct sources of operational advantage. The first concerns \emph{reference-frame resources}. Processes that are not covariant under independent local symmetry actions $G\times G$ implicitly rely on symmetry-breaking structure, such as shared reference frames or externally fixed bases, whose establishment is itself a physical task requiring causal resources~\cite{bartlett2007,gour2008,marvian2014,giacomini2019a,giacomini2019b,delahamette2020}. Enforcing local-frame covariance removes this form of background assistance and restricts attention to symmetry-invariant relational degrees of freedom.

The second concerns \emph{control at the operational interface}. Apparent device-independent advantages arise if interface-excluded degrees of freedom are \emph{modulated by the settings} in a way that is not recorded at the interface, for instance through unrecorded conditioning or postselection~\cite{lloyd2011}, or more generally through physical mechanisms that implement hidden routing or selection over interface-excluded variables. Interface-admissible coarse-grained reporting (Definition~\ref{def:interface_admissible}) rules out this form of assistance by fixing which setting dependences are admissible at the declared interface: physical implementations may employ arbitrary dynamics, but reported distributions that rely on unrecorded setting-dependent modulation are classified as background-assisted.

Within this two-axis classification, operational advantages associated with indefinite causal order~\cite{chiribella2012,guerin2016,guerin2019} can be grouped according to whether they rely on symmetry-breaking reference-frame resources and/or unrecorded control beyond the declared operational interface. One broad class of implementations is \emph{background-assisted}, in which at least one such resource is present. A distinct class consists of genuinely \emph{background-free} implementations, in which the process is locally frame covariant, the reported data are interface-admissible, and any nonclassical causal structure is encoded entirely within symmetry-invariant relational degrees of freedom, specifically within non-CC multiplicity subsystems.

\emph{Background-free causal structures and ICO classifications.}
This perspective resolves the conceptual tension emphasized in the Introduction.
Establishing a shared reference frame is a concrete physical task that presupposes some causal organization between laboratories; treating such an alignment as an unexamined background resource therefore risks circularity when analyzing correlations without predefined causal order. In the present framework, background-free nonclassical causal signatures---when they exist---are not attributed to externally supplied alignment or hidden interface-level control, but to symmetry-invariant relational structure compatible with local-frame covariance and interface admissibility. As in the standard process-matrix approach, this relational structure is treated as an operational primitive---specified only through its observable input–output correlations and not through an assumed causal preparation---thereby avoiding hidden reliance on reference-frame establishment while remaining within the usual device-independent operational assumptions.

More broadly, this work contributes to ongoing efforts to classify forms of indefinite causal order by adding an explicitly operational and physically grounded dimension. Existing approaches organize causal structure according to abstract properties of process matrices, such as purification postulates~\cite{araujo2017}, compositional structure~\cite{bisio2019}, or no-signalling constraints~\cite{apadula2024}, or by modifying the underlying kinematic notion
of a process through alternative system decompositions, such as time-delocalized implementations
\cite{oreshkov2019,wechs2019,wechs2023,vilasini2024a,vilasini2024b}. The present framework is complementary: rather than refining the space of abstract processes or altering the kinematic substrate, it identifies which device-independent causal signatures remain certifiable once reference-frame background is removed and hidden interface-level control is excluded. In this way, it clarifies which manifestations of indefinite causal order reflect intrinsic symmetry-invariant relational structure and which rely on background assistance.

\emph{Beyond GYNI.}
Although our analysis was instantiated on the bipartite GYNI causal game, the underlying reasoning is not specific to this task. The introduction of symmetry-induced latent labels, the distinction between resolved and coarse-grained operational interfaces, and the criterion of
interface-admissible reporting (NC/IC) apply to device-independent causal scenarios in which physically meaningful degrees of freedom are excluded from the declared interface. Likewise, the identification of non-CC multiplicity as a necessary symmetry-invariant resource reflects a
structural property of covariant process implementations rather than a feature of a particular inequality. Extending the present analysis to other causal games or multipartite scenarios would require a task-specific specification of the operational interface and corresponding certification criteria, but does not alter the covariance principle or the interface-level admissibility logic developed here.

\emph{Limitations and scope.} Our analysis rests on a few key assumptions that define its scope. First, the no-go results apply specifically to a fixed coarse-grained operational interface, in which symmetry-induced labels are excluded from the reported data; enlarging the interface would define a different device-independent task to which the present analysis does not apply. Second, while non-classical--classical (non-CC) multiplicity is shown to be necessary for genuinely background-free violations, it is not generically sufficient, as realizing observable violations depends on additional task-dependent compatibility conditions and on satisfying the reduced fixed-point constraints on the invariant blocks. Finally, our analysis is confined to single-shot scenarios. It does not address how sequential or adaptive use of a covariant process over multiple rounds might exploit the underlying symmetry-invariant structure. In particular, adaptive strategies could activate relational quantum resources that are inaccessible at the single-shot interface. More generally, multi-round protocols may rely on additional physical resources—such as memory, calibration, or communication across rounds—that fall outside the background-free assumptions imposed here, and whose effect on device-independent causal certification is not captured by the present analysis.

\emph{Future directions.} Taken together, these results indicate that indefinite causal order is best understood as a relational phenomenon whose device-independent certification is tightly constrained by symmetry, operational interface choice, and the internal structure of relational degrees of freedom. Clarifying how such relational resources can be accessed, combined, or amplified across interfaces and protocols remains an open direction for future work.

\bibliographystyle{apsrev4-2}
\bibliography{main}

\begin{thebibliography}{24}%
\makeatletter
\providecommand \@ifxundefined [1]{%
 \@ifx{#1\undefined}
}%
\providecommand \@ifnum [1]{%
 \ifnum #1\expandafter \@firstoftwo
 \else \expandafter \@secondoftwo
 \fi
}%
\providecommand \@ifx [1]{%
 \ifx #1\expandafter \@firstoftwo
 \else \expandafter \@secondoftwo
 \fi
}%
\providecommand \natexlab [1]{#1}%
\providecommand \enquote  [1]{``#1''}%
\providecommand \bibnamefont  [1]{#1}%
\providecommand \bibfnamefont [1]{#1}%
\providecommand \citenamefont [1]{#1}%
\providecommand \href@noop [0]{\@secondoftwo}%
\providecommand \href [0]{\begingroup \@sanitize@url \@href}%
\providecommand \@href[1]{\@@startlink{#1}\@@href}%
\providecommand \@@href[1]{\endgroup#1\@@endlink}%
\providecommand \@sanitize@url [0]{\catcode `\\12\catcode `\$12\catcode `\&12\catcode `\#12\catcode `\^12\catcode `\_12\catcode `\%12\relax}%
\providecommand \@@startlink[1]{}%
\providecommand \@@endlink[0]{}%
\providecommand \url  [0]{\begingroup\@sanitize@url \@url }%
\providecommand \@url [1]{\endgroup\@href {#1}{\urlprefix }}%
\providecommand \urlprefix  [0]{URL }%
\providecommand \Eprint [0]{\href }%
\providecommand \doibase [0]{https://doi.org/}%
\providecommand \selectlanguage [0]{\@gobble}%
\providecommand \bibinfo  [0]{\@secondoftwo}%
\providecommand \bibfield  [0]{\@secondoftwo}%
\providecommand \translation [1]{[#1]}%
\providecommand \BibitemOpen [0]{}%
\providecommand \bibitemStop [0]{}%
\providecommand \bibitemNoStop [0]{.\EOS\space}%
\providecommand \EOS [0]{\spacefactor3000\relax}%
\providecommand \BibitemShut  [1]{\csname bibitem#1\endcsname}%
\let\auto@bib@innerbib\@empty
\bibitem [{\citenamefont {Oreshkov}\ \emph {et~al.}(2012)\citenamefont {Oreshkov}, \citenamefont {Costa},\ and\ \citenamefont {Brukner}}]{oreshkov2012}%
  \BibitemOpen
  \bibfield  {author} {\bibinfo {author} {\bibfnamefont {O.}~\bibnamefont {Oreshkov}}, \bibinfo {author} {\bibfnamefont {F.}~\bibnamefont {Costa}},\ and\ \bibinfo {author} {\bibfnamefont {{\v C}.}~\bibnamefont {Brukner}},\ }\href {https://doi.org/10.1038/ncomms2076} {\bibfield  {journal} {\bibinfo  {journal} {Nature Communications}\ }\textbf {\bibinfo {volume} {3}},\ \bibinfo {pages} {1092} (\bibinfo {year} {2012})}\BibitemShut {NoStop}%
\bibitem [{\citenamefont {Chiribella}\ \emph {et~al.}(2013)\citenamefont {Chiribella}, \citenamefont {D'Ariano}, \citenamefont {Perinotti},\ and\ \citenamefont {Valiron}}]{chiribella2013}%
  \BibitemOpen
  \bibfield  {author} {\bibinfo {author} {\bibfnamefont {G.}~\bibnamefont {Chiribella}}, \bibinfo {author} {\bibfnamefont {G.~M.}\ \bibnamefont {D'Ariano}}, \bibinfo {author} {\bibfnamefont {P.}~\bibnamefont {Perinotti}},\ and\ \bibinfo {author} {\bibfnamefont {B.}~\bibnamefont {Valiron}},\ }\href {https://doi.org/10.1103/PhysRevA.88.022318} {\bibfield  {journal} {\bibinfo  {journal} {Phys. Rev. A}\ }\textbf {\bibinfo {volume} {88}},\ \bibinfo {pages} {022318} (\bibinfo {year} {2013})}\BibitemShut {NoStop}%
\bibitem [{\citenamefont {Brukner}(2014)}]{brukner2014}%
  \BibitemOpen
  \bibfield  {author} {\bibinfo {author} {\bibfnamefont {{\v C}.}~\bibnamefont {Brukner}},\ }\href {https://doi.org/10.1038/nphys2930} {\bibfield  {journal} {\bibinfo  {journal} {Nature Physics}\ }\textbf {\bibinfo {volume} {10}},\ \bibinfo {pages} {259} (\bibinfo {year} {2014})}\BibitemShut {NoStop}%
\bibitem [{\citenamefont {Bartlett}\ \emph {et~al.}(2007)\citenamefont {Bartlett}, \citenamefont {Rudolph},\ and\ \citenamefont {Spekkens}}]{bartlett2007}%
  \BibitemOpen
  \bibfield  {author} {\bibinfo {author} {\bibfnamefont {S.~D.}\ \bibnamefont {Bartlett}}, \bibinfo {author} {\bibfnamefont {T.}~\bibnamefont {Rudolph}},\ and\ \bibinfo {author} {\bibfnamefont {R.~W.}\ \bibnamefont {Spekkens}},\ }\href {https://doi.org/10.1103/RevModPhys.79.555} {\bibfield  {journal} {\bibinfo  {journal} {Reviews of Modern Physics}\ }\textbf {\bibinfo {volume} {79}},\ \bibinfo {pages} {555} (\bibinfo {year} {2007})}\BibitemShut {NoStop}%
\bibitem [{\citenamefont {Gour}\ and\ \citenamefont {Spekkens}(2008)}]{gour2008}%
  \BibitemOpen
  \bibfield  {author} {\bibinfo {author} {\bibfnamefont {G.}~\bibnamefont {Gour}}\ and\ \bibinfo {author} {\bibfnamefont {R.~W.}\ \bibnamefont {Spekkens}},\ }\href {https://doi.org/10.1088/1367-2630/10/3/033023} {\bibfield  {journal} {\bibinfo  {journal} {New Journal of Physics}\ }\textbf {\bibinfo {volume} {10}},\ \bibinfo {pages} {033023} (\bibinfo {year} {2008})}\BibitemShut {NoStop}%
\bibitem [{\citenamefont {Marvian}\ and\ \citenamefont {Spekkens}(2014)}]{marvian2014}%
  \BibitemOpen
  \bibfield  {author} {\bibinfo {author} {\bibfnamefont {I.}~\bibnamefont {Marvian}}\ and\ \bibinfo {author} {\bibfnamefont {R.~W.}\ \bibnamefont {Spekkens}},\ }\href {https://doi.org/10.1038/ncomms4821} {\bibfield  {journal} {\bibinfo  {journal} {Nature Communications}\ }\textbf {\bibinfo {volume} {5}},\ \bibinfo {pages} {3821} (\bibinfo {year} {2014})}\BibitemShut {NoStop}%
\bibitem [{\citenamefont {Giacomini}\ \emph {et~al.}(2019{\natexlab{a}})\citenamefont {Giacomini}, \citenamefont {Castro-Ruiz},\ and\ \citenamefont {Brukner}}]{giacomini2019a}%
  \BibitemOpen
  \bibfield  {author} {\bibinfo {author} {\bibfnamefont {F.}~\bibnamefont {Giacomini}}, \bibinfo {author} {\bibfnamefont {E.}~\bibnamefont {Castro-Ruiz}},\ and\ \bibinfo {author} {\bibfnamefont {{\v{C}}.}~\bibnamefont {Brukner}},\ }\href {https://doi.org/10.1038/s41467-018-08155-0} {\bibfield  {journal} {\bibinfo  {journal} {Nature Communications}\ }\textbf {\bibinfo {volume} {10}},\ \bibinfo {pages} {494} (\bibinfo {year} {2019}{\natexlab{a}})}\BibitemShut {NoStop}%
\bibitem [{\citenamefont {Giacomini}\ \emph {et~al.}(2019{\natexlab{b}})\citenamefont {Giacomini}, \citenamefont {Castro-Ruiz},\ and\ \citenamefont {Brukner}}]{giacomini2019b}%
  \BibitemOpen
  \bibfield  {author} {\bibinfo {author} {\bibfnamefont {F.}~\bibnamefont {Giacomini}}, \bibinfo {author} {\bibfnamefont {E.}~\bibnamefont {Castro-Ruiz}},\ and\ \bibinfo {author} {\bibfnamefont {{\v{C}}.}~\bibnamefont {Brukner}},\ }\href {https://doi.org/10.1103/PhysRevLett.123.090404} {\bibfield  {journal} {\bibinfo  {journal} {Physical Review Letters}\ }\textbf {\bibinfo {volume} {123}},\ \bibinfo {pages} {090404} (\bibinfo {year} {2019}{\natexlab{b}})}\BibitemShut {NoStop}%
\bibitem [{\citenamefont {de~la Hamette}\ and\ \citenamefont {Galley}(2020)}]{delahamette2020}%
  \BibitemOpen
  \bibfield  {author} {\bibinfo {author} {\bibfnamefont {A.-C.}\ \bibnamefont {de~la Hamette}}\ and\ \bibinfo {author} {\bibfnamefont {T.~D.}\ \bibnamefont {Galley}},\ }\href {https://doi.org/10.22331/q-2020-11-30-367} {\bibfield  {journal} {\bibinfo  {journal} {Quantum}\ }\textbf {\bibinfo {volume} {4}},\ \bibinfo {pages} {367} (\bibinfo {year} {2020})}\BibitemShut {NoStop}%
\bibitem [{\citenamefont {Parker}\ and\ \citenamefont {Costa}(2022)}]{parker2022}%
  \BibitemOpen
  \bibfield  {author} {\bibinfo {author} {\bibfnamefont {L.}~\bibnamefont {Parker}}\ and\ \bibinfo {author} {\bibfnamefont {F.}~\bibnamefont {Costa}},\ }\href {https://doi.org/10.22331/q-2022-11-28-865} {\bibfield  {journal} {\bibinfo  {journal} {Quantum}\ }\textbf {\bibinfo {volume} {6}},\ \bibinfo {pages} {865} (\bibinfo {year} {2022})}\BibitemShut {NoStop}%
\bibitem [{\citenamefont {Ara\'ujo}\ \emph {et~al.}(2015)\citenamefont {Ara\'ujo}, \citenamefont {Branciard}, \citenamefont {Costa}, \citenamefont {Feix}, \citenamefont {Giarmatzi},\ and\ \citenamefont {Brukner}}]{araujo2015}%
  \BibitemOpen
  \bibfield  {author} {\bibinfo {author} {\bibfnamefont {M.}~\bibnamefont {Ara\'ujo}}, \bibinfo {author} {\bibfnamefont {C.}~\bibnamefont {Branciard}}, \bibinfo {author} {\bibfnamefont {F.}~\bibnamefont {Costa}}, \bibinfo {author} {\bibfnamefont {A.}~\bibnamefont {Feix}}, \bibinfo {author} {\bibfnamefont {C.}~\bibnamefont {Giarmatzi}},\ and\ \bibinfo {author} {\bibfnamefont {{\v C}.}~\bibnamefont {Brukner}},\ }\href {https://doi.org/10.1088/1367-2630/17/10/102001} {\bibfield  {journal} {\bibinfo  {journal} {New Journal of Physics}\ }\textbf {\bibinfo {volume} {17}},\ \bibinfo {pages} {102001} (\bibinfo {year} {2015})}\BibitemShut {NoStop}%
\bibitem [{\citenamefont {Branciard}\ \emph {et~al.}(2015)\citenamefont {Branciard}, \citenamefont {Ara\'ujo}, \citenamefont {Feix}, \citenamefont {Costa},\ and\ \citenamefont {Brukner}}]{branciard2016}%
  \BibitemOpen
  \bibfield  {author} {\bibinfo {author} {\bibfnamefont {C.}~\bibnamefont {Branciard}}, \bibinfo {author} {\bibfnamefont {M.}~\bibnamefont {Ara\'ujo}}, \bibinfo {author} {\bibfnamefont {A.}~\bibnamefont {Feix}}, \bibinfo {author} {\bibfnamefont {F.}~\bibnamefont {Costa}},\ and\ \bibinfo {author} {\bibfnamefont {{\v C}.}~\bibnamefont {Brukner}},\ }\href {https://doi.org/10.1088/1367-2630/18/1/013008} {\bibfield  {journal} {\bibinfo  {journal} {New Journal of Physics}\ }\textbf {\bibinfo {volume} {18}},\ \bibinfo {pages} {013008} (\bibinfo {year} {2015})}\BibitemShut {NoStop}%
\bibitem [{\citenamefont {Lloyd}\ \emph {et~al.}(2011)\citenamefont {Lloyd}, \citenamefont {Maccone}, \citenamefont {Garcia-Patron}, \citenamefont {Giovannetti}, \citenamefont {Shikano}, \citenamefont {Pirandola}, \citenamefont {Rozema}, \citenamefont {Darabi}, \citenamefont {Soudagar}, \citenamefont {Shalm},\ and\ \citenamefont {Steinberg}}]{lloyd2011}%
  \BibitemOpen
  \bibfield  {author} {\bibinfo {author} {\bibfnamefont {S.}~\bibnamefont {Lloyd}}, \bibinfo {author} {\bibfnamefont {L.}~\bibnamefont {Maccone}}, \bibinfo {author} {\bibfnamefont {R.}~\bibnamefont {Garcia-Patron}}, \bibinfo {author} {\bibfnamefont {V.}~\bibnamefont {Giovannetti}}, \bibinfo {author} {\bibfnamefont {Y.}~\bibnamefont {Shikano}}, \bibinfo {author} {\bibfnamefont {S.}~\bibnamefont {Pirandola}}, \bibinfo {author} {\bibfnamefont {L.~A.}\ \bibnamefont {Rozema}}, \bibinfo {author} {\bibfnamefont {A.}~\bibnamefont {Darabi}}, \bibinfo {author} {\bibfnamefont {Y.}~\bibnamefont {Soudagar}}, \bibinfo {author} {\bibfnamefont {L.~K.}\ \bibnamefont {Shalm}},\ and\ \bibinfo {author} {\bibfnamefont {A.~M.}\ \bibnamefont {Steinberg}},\ }\href {https://doi.org/10.1103/PhysRevLett.106.040403} {\bibfield  {journal} {\bibinfo  {journal} {Phys. Rev. Lett.}\ }\textbf {\bibinfo {volume} {106}},\ \bibinfo {pages} {040403} (\bibinfo {year} {2011})}\BibitemShut {NoStop}%
\bibitem [{\citenamefont {Chiribella}(2012)}]{chiribella2012}%
  \BibitemOpen
  \bibfield  {author} {\bibinfo {author} {\bibfnamefont {G.}~\bibnamefont {Chiribella}},\ }\href {https://doi.org/10.1103/PhysRevA.86.040301} {\bibfield  {journal} {\bibinfo  {journal} {Phys. Rev. A}\ }\textbf {\bibinfo {volume} {86}},\ \bibinfo {pages} {040301} (\bibinfo {year} {2012})}\BibitemShut {NoStop}%
\bibitem [{\citenamefont {Gu\'erin}\ \emph {et~al.}(2016)\citenamefont {Gu\'erin}, \citenamefont {Feix}, \citenamefont {Ara\'ujo},\ and\ \citenamefont {Brukner}}]{guerin2016}%
  \BibitemOpen
  \bibfield  {author} {\bibinfo {author} {\bibfnamefont {P.~A.}\ \bibnamefont {Gu\'erin}}, \bibinfo {author} {\bibfnamefont {A.}~\bibnamefont {Feix}}, \bibinfo {author} {\bibfnamefont {M.}~\bibnamefont {Ara\'ujo}},\ and\ \bibinfo {author} {\bibfnamefont {{\v C}.}~\bibnamefont {Brukner}},\ }\href {https://doi.org/10.1103/PhysRevLett.117.100502} {\bibfield  {journal} {\bibinfo  {journal} {Physical Review Letters}\ }\textbf {\bibinfo {volume} {117}},\ \bibinfo {pages} {100502} (\bibinfo {year} {2016})}\BibitemShut {NoStop}%
\bibitem [{\citenamefont {Gu\'erin}\ \emph {et~al.}(2019)\citenamefont {Gu\'erin}, \citenamefont {Rubino},\ and\ \citenamefont {Brukner}}]{guerin2019}%
  \BibitemOpen
  \bibfield  {author} {\bibinfo {author} {\bibfnamefont {P.~A.}\ \bibnamefont {Gu\'erin}}, \bibinfo {author} {\bibfnamefont {G.}~\bibnamefont {Rubino}},\ and\ \bibinfo {author} {\bibfnamefont {{\v C}.}~\bibnamefont {Brukner}},\ }\href {https://doi.org/10.1103/PhysRevA.99.062317} {\bibfield  {journal} {\bibinfo  {journal} {Phys. Rev. A}\ }\textbf {\bibinfo {volume} {99}},\ \bibinfo {pages} {062317} (\bibinfo {year} {2019})}\BibitemShut {NoStop}%
\bibitem [{\citenamefont {Ara\'ujo}\ \emph {et~al.}(2017)\citenamefont {Ara\'ujo}, \citenamefont {Feix}, \citenamefont {Navascu\'es},\ and\ \citenamefont {Brukner}}]{araujo2017}%
  \BibitemOpen
  \bibfield  {author} {\bibinfo {author} {\bibfnamefont {M.}~\bibnamefont {Ara\'ujo}}, \bibinfo {author} {\bibfnamefont {A.}~\bibnamefont {Feix}}, \bibinfo {author} {\bibfnamefont {M.}~\bibnamefont {Navascu\'es}},\ and\ \bibinfo {author} {\bibfnamefont {{\v C}.}~\bibnamefont {Brukner}},\ }\href {https://doi.org/10.22331/q-2017-04-26-10} {\bibfield  {journal} {\bibinfo  {journal} {Quantum}\ }\textbf {\bibinfo {volume} {1}},\ \bibinfo {pages} {10} (\bibinfo {year} {2017})}\BibitemShut {NoStop}%
\bibitem [{\citenamefont {Bisio}\ and\ \citenamefont {Perinotti}(2019)}]{bisio2019}%
  \BibitemOpen
  \bibfield  {author} {\bibinfo {author} {\bibfnamefont {A.}~\bibnamefont {Bisio}}\ and\ \bibinfo {author} {\bibfnamefont {P.}~\bibnamefont {Perinotti}},\ }\href {https://doi.org/10.1098/rspa.2018.0706} {\bibfield  {journal} {\bibinfo  {journal} {Proceedings of the Royal Society A}\ }\textbf {\bibinfo {volume} {475}},\ \bibinfo {pages} {20180706} (\bibinfo {year} {2019})}\BibitemShut {NoStop}%
\bibitem [{\citenamefont {Apadula}\ \emph {et~al.}(2024)\citenamefont {Apadula}, \citenamefont {Bisio},\ and\ \citenamefont {Perinotti}}]{apadula2024}%
  \BibitemOpen
  \bibfield  {author} {\bibinfo {author} {\bibfnamefont {L.}~\bibnamefont {Apadula}}, \bibinfo {author} {\bibfnamefont {A.}~\bibnamefont {Bisio}},\ and\ \bibinfo {author} {\bibfnamefont {P.}~\bibnamefont {Perinotti}},\ }\href {https://doi.org/10.22331/q-2024-02-05-1241} {\bibfield  {journal} {\bibinfo  {journal} {Quantum}\ }\textbf {\bibinfo {volume} {8}},\ \bibinfo {pages} {1241} (\bibinfo {year} {2024})}\BibitemShut {NoStop}%
\bibitem [{\citenamefont {Oreshkov}(2019)}]{oreshkov2019}%
  \BibitemOpen
  \bibfield  {author} {\bibinfo {author} {\bibfnamefont {O.}~\bibnamefont {Oreshkov}},\ }\href {https://doi.org/10.22331/q-2019-12-02-206} {\bibfield  {journal} {\bibinfo  {journal} {Quantum}\ }\textbf {\bibinfo {volume} {3}},\ \bibinfo {pages} {206} (\bibinfo {year} {2019})}\BibitemShut {NoStop}%
\bibitem [{\citenamefont {Wechs}\ \emph {et~al.}(2019)\citenamefont {Wechs}, \citenamefont {Abbott},\ and\ \citenamefont {Branciard}}]{wechs2019}%
  \BibitemOpen
  \bibfield  {author} {\bibinfo {author} {\bibfnamefont {J.}~\bibnamefont {Wechs}}, \bibinfo {author} {\bibfnamefont {A.~A.}\ \bibnamefont {Abbott}},\ and\ \bibinfo {author} {\bibfnamefont {C.}~\bibnamefont {Branciard}},\ }\href {https://doi.org/10.1088/1367-2630/aaf352} {\bibfield  {journal} {\bibinfo  {journal} {New Journal of Physics}\ }\textbf {\bibinfo {volume} {21}},\ \bibinfo {pages} {013027} (\bibinfo {year} {2019})}\BibitemShut {NoStop}%
\bibitem [{\citenamefont {Wechs}\ \emph {et~al.}(2023)\citenamefont {Wechs}, \citenamefont {Branciard},\ and\ \citenamefont {Oreshkov}}]{wechs2023}%
  \BibitemOpen
  \bibfield  {author} {\bibinfo {author} {\bibfnamefont {J.}~\bibnamefont {Wechs}}, \bibinfo {author} {\bibfnamefont {C.}~\bibnamefont {Branciard}},\ and\ \bibinfo {author} {\bibfnamefont {O.}~\bibnamefont {Oreshkov}},\ }\href {https://doi.org/10.1038/s41467-023-36893-3} {\bibfield  {journal} {\bibinfo  {journal} {Nature Communications}\ }\textbf {\bibinfo {volume} {14}},\ \bibinfo {pages} {1471} (\bibinfo {year} {2023})}\BibitemShut {NoStop}%
\bibitem [{\citenamefont {Vilasini}\ and\ \citenamefont {Renner}(2024{\natexlab{a}})}]{vilasini2024a}%
  \BibitemOpen
  \bibfield  {author} {\bibinfo {author} {\bibfnamefont {V.}~\bibnamefont {Vilasini}}\ and\ \bibinfo {author} {\bibfnamefont {R.}~\bibnamefont {Renner}},\ }\href {https://doi.org/10.1103/PhysRevA.110.022227} {\bibfield  {journal} {\bibinfo  {journal} {Physical Review A}\ }\textbf {\bibinfo {volume} {110}},\ \bibinfo {pages} {022227} (\bibinfo {year} {2024}{\natexlab{a}})}\BibitemShut {NoStop}%
\bibitem [{\citenamefont {Vilasini}\ and\ \citenamefont {Renner}(2024{\natexlab{b}})}]{vilasini2024b}%
  \BibitemOpen
  \bibfield  {author} {\bibinfo {author} {\bibfnamefont {V.}~\bibnamefont {Vilasini}}\ and\ \bibinfo {author} {\bibfnamefont {R.}~\bibnamefont {Renner}},\ }\href {https://doi.org/10.1103/PhysRevLett.133.080201} {\bibfield  {journal} {\bibinfo  {journal} {Phys. Rev. Lett.}\ }\textbf {\bibinfo {volume} {133}},\ \bibinfo {pages} {080201} (\bibinfo {year} {2024}{\natexlab{b}})}\BibitemShut {NoStop}%
\end{thebibliography}%

\end{document}